# The Debye temperature of σ-phase Fe-Mo compounds as determined with Mössbauer spectroscopy

J. Cieslak and S. M. Dubiel[*]

AGH University of Science and Technology, Faculty of Physics and Applied Computer Science, 30-059 Krakow, Poland

**Abstract**

The Debye temperature, $\Theta_D$, of σ-phase $Fe_{100-x}Mo_x$ compounds with $47 \leq x \leq 56.7$ was determined from the temperature dependence of the centre shift of Mössbauer spectra recorded in the temperature range of 80 – 300 K. Its compositional dependence shows a weak increase with $x$ whose rate, in a linear approximation, is equal to 3.1 K/at%. The results are compared with the corresponding ones found previously for the σ-phase in Fe-Cr and Fe-V compounds.

[*] Corresponding author: dubiel@novell.ftj.agh.edu.pl (S. M. Dubiel)

The σ-phase in the Fe-Mo alloy system is one of numerous members of the so-called Frank-Kasper phases, which are topologically close-packed structures. Their most characteristic feature is a complex crystallographic structure and high coordination numbers (CN). In the case of σ, a tetragonal unit cell hosts 30 atoms distributed over five non-equivalent lattice sites with CN = 12 – 15 [1]. Among about 50 known examples of σ revealed so far to exist in a binary alloy systems, its physical properties were systematically studied only in two systems viz. Fe-Cr and Fe-V [2]. In the case of σ-FeMo compounds, papers available in the literature were, to our best knowledge, exclusively devoted to determining its crystallographic structure and ordering of atoms over the lattice sites [1,3-6] while its physical properties still remain to be discovered. To fill this gap, we are reporting here results on the Debye temperature, $\Theta_D$, that we obtained on a series of σ-$Fe_{100-x}Mo_x$ compounds ($47 \leq x \leq 56.7$) using the Mössbauer spectroscopy (MS). It is well known that the actual value of $\Theta_D$ of a given solid significantly depends on a method used for its determination. Consequently, a comparison of $\Theta_D$ – values determined with different experimental techniques may be misleading. For example $\Theta_D$ of iron equals to 418 K as found from the X-ray diffraction experiment [7], and to 477 K as estimated from the elastic constant [8]. On the other hand, a comparison of the results obtained for different solids with the same method seems reasonable as it may deliver useful information on the lattice dynamics of the solids. Using MS technique we have already measured $\Theta_D$ – from a temperature dependence of the centre shift, *CS*, - for the σ-phase in Fe-Cr [9] and Fe-V [10] compounds, revealing a linear compositional dependence in the former case and a non-monotonous one in the latter.

Seven samples of σ-$Fe_{100-x}Mo_x$ (x=47, 48, 49, 51, 53, 55 and 56.7) were produced by a sintering method. For this purpose, powdered iron (99.9% purity) and molybdenum (99.95 % purity) were mixed in appropriate proportions and next compressed into pellets. The pellets were annealed in vacuum at 1700 K for 6 hours and afterwards quenched into liquid nitrogen, which resulted in their transformation into the σ-phase as revealed by the X-ray diffraction patterns recorded on the samples after the sintering process was over.

To get the quantity of merit for determining $\Theta_D$, a series of Mössbauer spectra was recorded in a transmission geometry for each sample in the temperature range of 80 – 300 K using a standard spectrometer and a $^{57}$Co/Rh source of 14.4 keV gamma rays. Temperature of the samples, which were kept in a cryostat, was stabilized with an accuracy of ± 0.2 K. Examples of the recorded spectra are shown in Fig. 1a. Due to a lack of a visible structure in the spectra, they were fitted in terms of the isomer shift distribution method [11] that yielded the isomer shift distribution (ISD) curves. Those corresponding to the spectra presented in Fig. 1a are displayed in Fig. 1b. By integration of ISD-curves, the average centre shift, *<CS>*, was determined. Its temperature dependence can be expressed as follows:

$$<CS>(T) = IS(T) + SODS(T) \qquad (1)$$

Where *IS (T)* is the isomer shift, a spectral quantity related to a charge density at the probe nucleus. As discussed elsewhere [12], its temperature dependence is weak, so it is usually approximated by a constant term, *IS(0) that* is eventually composition dependent. *SODS* term is the so-called second-order Doppler shift. It shows strong temperature dependence. Assuming the whole temperature dependence of *<CS>* goes via *SODS* term and using the Debye model for the phonon spectrum one arrives at the following formula:

$$<CS(T)> = IS(0) - \frac{3kT}{2mc}\left[\frac{3\Theta_D}{8T} + 3\left(\frac{T}{\Theta_D}\right)^3 \int_0^{\Theta_D/T} \frac{x^3}{e^x-1}dx\right] \qquad (2)$$

Where $m$ is the mass of the $^{57}$Fe nucleus, $k$ is the Boltzmann constant; $c$ is the velocity of light.

Fitting equation (2) to the $<CS>(T)$ – values, determined by the applied fitting procedure (a typical temperature dependence of $<CS>$ is illustrated in Fig. 2), enabled determination of the $\Theta_D$ - values. They can be seen in Fig. 3 where, for comparison, the corresponding data obtained previously for the σ-phase in Fe-Cr and Fe-V compounds are added. The $\Theta_D$ – values exhibit a weak increase with $x$ which, in a linear approximation, can be described as $\Theta_D = 257.25 + 3.14\ x$. Figure 3 gives a clear evidence that the Debye temperature is characteristic of a given alloy system, and the $\Theta_D$ – values found for Fe-Mo are the lowest ones among the three systems. It is also obvious that the effect of Cr, V and Mo on $\Theta_D$ – values in the three alloy systems has nothing to do with the Debye temperature of chromium (630 K), vanadium (380 K) nor molybdenum (450 K) – the values obtained from the specific heat measurements [13].

**Acknowledgement**
The results presented in this paper were obtained in the frame of the project supported by The Ministry of Science and Higher Education, Warsaw (grant No. N N202 228837).

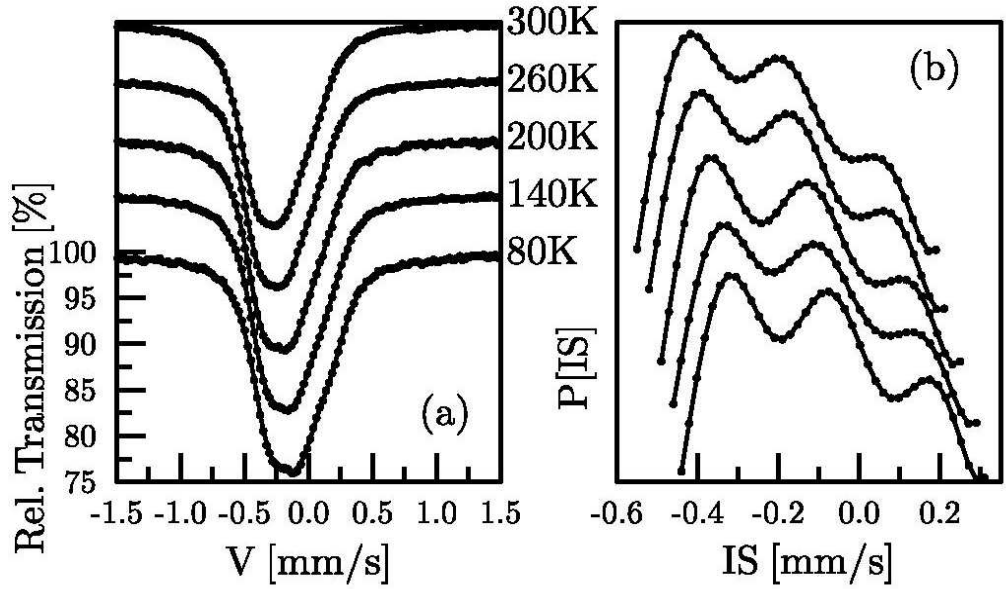

**Fig. 1:** (a) Room temperature Mössbauer spectra recorded on the σ-$Fe_{47}Mo_{53}$ sample at various temperatures shown, and (b) the corresponding ISD-curves. The solid lines are the best-fit to the experimental data in terms of the applied fitting procedure.

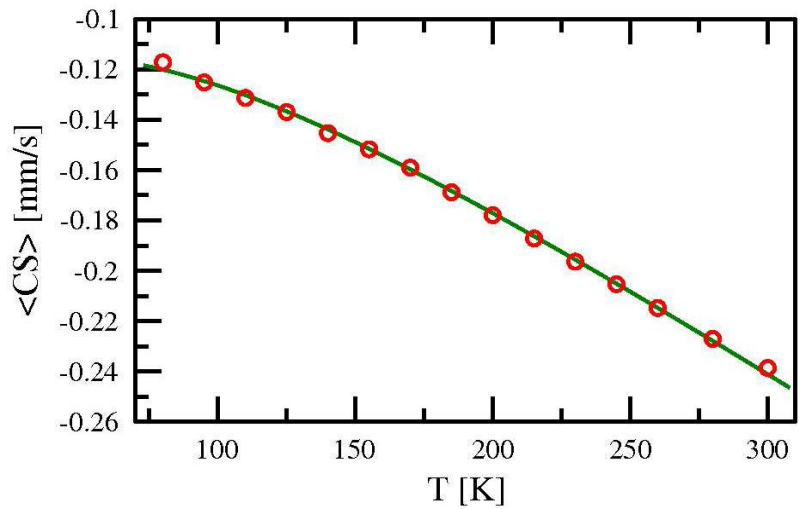

**Fig. 2:** Temperature dependence of the average centre shift, $<CS>$, as found for the σ-$Fe_{47}Mo_{53}$ compound. The solid line represents the best-fit to the experimental data in terms of equation (2).

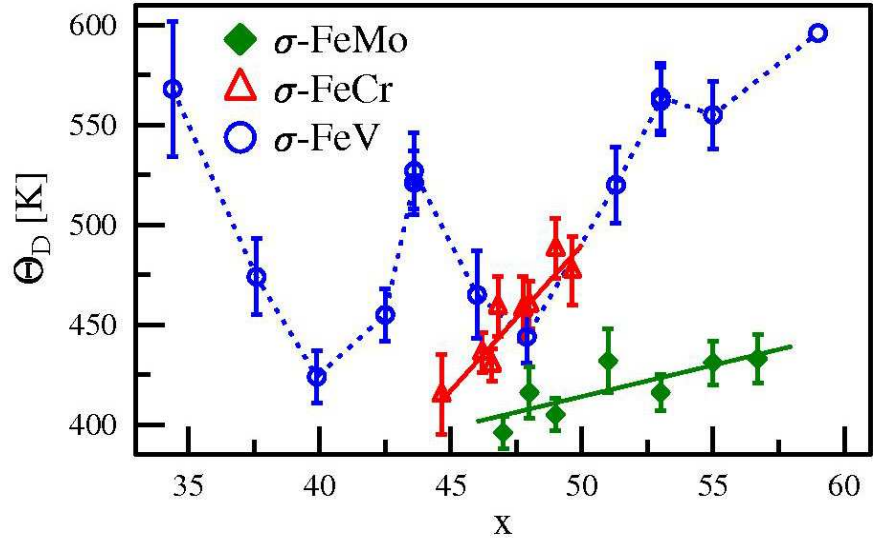

**Fig. 3:** Debye temperature, $\Theta_D$, versus *x* for the investigated samples. Previously found data for σ-$Fe_{100-x}Cr_x$ [9] and σ-$Fe_{100-x}V_x$ [10] compounds have been added for comparison. Straight lines represent the best linear fits to the data while the dotted line is to guide the eye.